\documentclass[11pt,twoside,a4paper,paper,notoc]{JHEP}
\usepackage{latexsym}
\usepackage{amsmath}
\usepackage{graphics}
\usepackage{epsfig}
\usepackage{array}
\usepackage[latin1]{inputenc}
\usepackage{feynmp}
\newcommand{\topl}[1]{\ensuremath{\mbox{\textsf{\scriptsize #1}}}}

\newcommand{\expt}[3]{\ensuremath{\langle\!\, \mathcal{Q}_{#1}^{#2}\,
    \!\rangle_{\raisebox{-1mm}{\mbox{\textsf{\hspace{-1.5mm} \tiny #3}}}}} }
\newcommand{\expen}[3]{\ensuremath{\langle\!\, \mathcal{Q}_{pen}^{(#2)#1}\,
    \!\rangle_{\raisebox{-1mm}{\mbox{\textsf{\hspace{-1.5mm} \tiny #3}}}}} }
\newcommand{\smsum}[1]{\ensuremath{\textstyle \sum_{\raisebox{-1.6mm}[.0ex][.5ex]
    {\scriptsize $\!\!\!\!\!#1\ $}}}}
\newcommand{\lbd}[1]{\ensuremath{\lambda_{bd}^{#1}}}
\newcommand{\lbs}[1]{\ensuremath{\lambda_{bs}^{#1}}}
\newcommand{\BdsJpsi}{\ensuremath{B_{d,s}\to J/\psi\ \! \eta\ \!}}
\newcommand{\BdJpsi}{\ensuremath{B_{d}\to J/\psi\ \! \eta\ \!}}
\newcommand{\BsJpsi}{\ensuremath{B_{s}\to J/\psi\ \! \eta\ \!}}

\newcommand{\Bdsetallp}{\ensuremath{B_{d,s}\to \eta\ \! \ell^+ \ell^-}}
\newcommand{\Bsetallp}{\ensuremath{B_{s}\to \eta\ \! \ell^+ \ell^-}}

\newcommand{\Bsetammp}{\ensuremath{B_{s}\to \eta\ \! \mu^+ \mu^-}}
\newcommand{\Bdetammp}{\ensuremath{B_{d}\to \eta\ \! \mu^+ \mu^-}}
\newcommand{\BdKllp}{\ensuremath{B_d\to K \ell^+ \ell^-}}
\newcommand{\BdsJpsiK}{\ensuremath{B_{d,s}\to J/\psi\ \! K}}
\newcommand{\BsJpsiK}{\ensuremath{B_{s}\to J/\psi\ \! K}}
\newcommand{\BdJpsiK}{\ensuremath{B_d\to J/\psi\ \! K}}
\newcommand{\BsJpsiKbar}{\ensuremath{B_d\to J/\psi\ \!
\overline{K}^0}}

\newcommand{\OPE}[1]{\ensuremath{\mathcal{Q}_{#1}}}
\newcommand{\braket}[3]{\ensuremath{\langle\ \! #1\!\ | #2 |\ \! #3\
\!\rangle}}
\newcommand{\epspq}{\ensuremath{\varepsilon_{pq}}}
\newcommand{\DeltaMt}[1]{\ensuremath{\Delta M_{#1}t}}
\renewcommand{\Re}{\mbox{Re}}
\renewcommand{\Im}{\mbox{Im}}
\title{Branching ratios for \BdsJpsi\ and \Bdsetallp, extracting $\gamma$
from \BdsJpsi, and possibilities for constraining $C_{10A}$ in semileptonic B
decays}
\author{P. Z. Skands
\\HEP department -- Niels Bohr Institute -- Blegdamsvej
17 -- DK-2100 Copenhagen Ø \\zeiler@nbi.dk}
\abstract{Estimates of the branching ratios for \BdsJpsi\ and
\Bdsetallp\ are obtained by SU(3) relation to \BdJpsiK\ and
\BdKllp, respectively, as functions of the $\eta_1-\eta_8$ mixing
parameter, $\theta_P$. Based on these estimates, a discussion of
the prospects for HERA-B, CDF-II, and ATLAS on these processes is
given. The $CP$ violation in \BdsJpsi\ is analyzed in depth and a
method to extract the angle $\gamma$ of the unitarity triangle is
discussed. Finally, a possible method to constrain the Wilson
coefficient $C_{10A}$ from measurements on semileptonic $B$
decays such as \BdKllp\ and \Bsetallp\ is proposed along with a
discussion of the prospects for future experiments and form factor
calculations to reach the precision required for this method to be
interesting} \preprint{hep-ph/0010115} 
\begin{document}
\section{Introduction}\label{intro}
One of the most important processes relevant to the study of $CP$
violation (see \cite{aftertop} for an excellent review of this
topic) is the well-known ``gold-plated mode'' $\BdJpsiK_S$.
Besides its usefulness in extracting $\sin(2\beta)$ (see e.g.\
\cite{BaBar}), it has relatively recently been argued that this
process can also be used to extract the angle $\gamma$ when
combined with its $U$-spin partner $\BsJpsiK_S$
 \cite{extractinggamma}.

In this paper, an alternative to the gold-plated mode for the
extraction of $\gamma$ is presented: \BdsJpsi. Due to the
inherent problems in adapting factorization to non-leptonic
decays we use the completely general and model independent method
of quark topologies to analyze the structure of \BdsJpsi,
obtaining a relation between the $B_d$ and $B_s$ amplitudes
through SU(3) flavour symmetry (the approximate symmetry of $u,d,s$). The
structure of the two processes is such that the
phase, $e^{i\gamma}$, is CKM suppressed in \BsJpsi\ relative to
\BdJpsi, and so effects of $CP$ violation will be more easily
visible in the \BdJpsi\ decay. However, the extraction of
$\gamma$ from this process is plagued by the appearance of a
normalization factor which cannot be determined directly. Through
$U$ spin relation, the $CP$ averaged rates of \BsJpsi\ and
$\BdJpsi$ can be combined to fix this normalization. Thus,
$\gamma$ can be determined with a theoretical uncertainty
depending only on SU(3) breaking corrections and the
$\eta_1-\eta_8$ mixing angle, $\theta_P$. Furthermore, the $B_d$
amplitude itself is suppressed relative to the $B_s$ amplitude,
the inverse of the situation in the $\BdsJpsiK_S$ case. The
difference in production rates of $B_d$ and $B_s$ mesons in the
experiment compensates to some extent for this difference in the
case \BdsJpsi\ whereas it worsens the situation for $\BdsJpsiK_S$.

As an addendum, a method to constrain $C_{10A}$ is presented. In
semileptonic decays, $C_{10A}$ describes the effective coupling
of the axial OPE operator $\mathcal{O}_{10A}$ \cite{aftertop}. In
models of new physics, its value generally deviates from the SM
prediction, due to virtual particle contributions from New
Physics particles present in such models. In this paper, a method
is proposed which allows the elimination of large hadronic
uncertainties caused by the presence of intermediate $\psi$
resonances by measuring distributions in semileptonic decays and
using relations between them. It is, however, doubtful whether
this method can find immediate application due to the high
precision required both for theoretical and experimental input.

The outline of the paper is as follows: In section \ref{sec:qt},
we use the method of quark topologies to analyze the CKM structure
of the contributions to the \BdsJpsi\ decay amplitude (see
\cite{non-leptonic} for a recent update on this method). Noting
the Zweig and SU(3) suppressions of certain topologies and
assuming SU(3) symmetry of the strong interaction dynamics, a
simple relation can be obtained between the \BdJpsi\ and \BsJpsi\
amplitudes. In section \ref{sec:extract}, the procedure proposed
in \cite{extractinggamma} for extracting $\gamma$ from $\BdsJpsiK_S$
is adapted for \BdsJpsi, and the amplitude relation obtained in
the previous section is used to obtain the normalization of the
$CP$ averaged \BdJpsi\ rate.

In Section \ref{sec:brbjpsi}, a simplified picture of the
\BdsJpsi\ process is adopted and SU(3) symmetry is invoked to
obtain a relation between the amplitudes of \BdJpsi, \BsJpsi,
$\BsJpsiK_S$ to the already measured $\BdJpsiK_S$. In section
\ref{sec:bretallp}, essentially the same is employed to obtain
$B_{d,s}\to\eta$ form factors from those for
 $B_d\to K$ calculated by Light Cone Sum Rules (LCSR) in
\cite{ball98}\cite{article}\footnote{For a more recent
calculation, see \cite{khodjam00}.}. Using these, an estimate for
the \Bdsetallp\ branching ratios can be obtained. Both the
\BdsJpsi\ and the \Bdsetallp branchings depend on the degree of
octet-singlet mixing in the $\eta$ system, expressed through the
mixing angle, $\theta_P$.

In section \ref{sec:constrainc10}, a simple method is proposed to
constrain $C_{10A}$, the axial semileptonic Wilson coefficient.
We replace the theoretically poorly known quantity
$C_{9V}^{\mbox{eff}}$ in the $B_d\to K\tau^+\tau^-$ amplitude by
measurable decay distributions for the $B_d\to K\mu^+\mu^-$
process, yielding $C_{10A}$ as a function of the total branching
to $K\tau\tau$, the differential branching to $K\mu\mu$,
$|V_{ts}^* V_{tb}|^2$, and the form factors $f_+$ and $f_-$.
Estimates in various SUGRA models and the 2 Higgs doublet model
are considered, and the application of the same  procedure to the
case \Bsetallp\ is considered. Concluding remarks and outlook are
given in section \ref{sec:conclall}.


\section{Analysis of \BdsJpsi}
\label{sec:qt} The time-independent transition amplitudes for
$B^0$ and $\overline{B}^0$ states into the final $CP$ eigenstate,
$f_{CP}$, are parametrized as in \cite{extractinggamma}:
\begin{eqnarray}
\braket{f_{CP}}{H}{B^0_q} \equiv \mathcal{A}_q & = & N_q
\left[1-{a}_qe^{i{\theta}_q}e^{i\gamma}\right] \equiv N_q z_q
\label{eq:Adef}\\ \braket{f_{CP}}{H}{\overline{B}^0_q} \equiv
\overline{\mathcal{A}}_q & = & \eta N_q
\left[1-{a}_qe^{i{\theta}_q}e^{-i\gamma}\right] \equiv \eta
N_q\overline{z}_q \label{eq:Abardef}
\end{eqnarray}
where $\eta$ is the $CP$ eigenvalue of $f_{CP}$, $\theta$ is a
strong phase, and $q\in\{d,s\}$. The amplitude at time $t$ for an
initial $B/\overline{B}$ meson to decay then becomes:
\begin{eqnarray}
|A_q(t)|^2 & = & \frac{|N_q|^2}2\left[ (R_L+\epspq
R_L^1)e^{-\Gamma_Lt}
                 +(R_H+\epspq R_H^1)e^{-\Gamma_Ht}\right. \nonumber \\ & &
                 \left.\hspace{5.mm}+2e^{-\Gamma t}\left[(A_D+\epspq
                 A_D^1)\cos(\DeltaMt{}) + (A_M(1-\frac{\epspq}{2})\sin(\DeltaMt{})
                 \right] \right] \label{eq:sqampt}\\ & & \nonumber \\
|\overline{A}_q(t)|^2 &
                 = & \frac{|N_q|^2}2\left[ (R_L+\epspq
                 \overline{R}_L^1)e^{-\Gamma_Lt} +(R_H+\epspq
                 \overline{R}_H^1)e^{-\Gamma_Ht}\right. \nonumber \\ & &
                 \left.\hspace{5.mm}-2e^{-\Gamma t}\left[(A_D+\epspq
                 \overline{A}_D^1)\cos(\DeltaMt{}) + (A_M(1+\frac{\epspq}{2})
                 \sin(\DeltaMt{}) \right] \right]\label{eq:sqamptbar}
\end{eqnarray}
where we define rate functions as in \cite{extractinggamma},
except that we here give the formulae to first order in the small
parameter $\epspq\equiv |\frac{p}{q}|^2-1$, where $p$ and $q$ are
the standard factors parametrizing the $B$ meson mass eigenstates
in terms of flavour states \cite{aftertop}. As there is, however,
only small possibility for new physics to result in a large
\epspq\, we henceforth ignore this parameter. The rate functions
entering the above are defined by ($\phi_q$ is the $B_q-\overline{B}_q$
mixing phase):
\begin{eqnarray}
R_L & \equiv & \frac12(|z_q|^2 +|\overline{z}_q|^2+2\eta \Re\{
             z_q^*\overline{z}_qe^{i\phi_q} \}) \label{def:RL} \\
R_H &
             \equiv & \frac12(|z_q|^2+|\overline{z}_q|^2-2\eta \Re\{
             z_q^*\overline{z}_qe^{i\phi_q} \}) \\
A_D & \equiv & \frac12
             (|z_q|^2-|\overline{z}_q|^2) \\
A_M & \equiv & -\eta\Im\{ z_q^*
             \overline{z}_q e^{i\phi_q} \}\label{def:AM} \\
R_L^1 & \equiv
             & -\frac12(|\overline{z}_q|^2+\eta \Re\{
             z_q^*\overline{z}_qe^{i\phi_q} \})  = -\frac12 (R_L+A_D)
             \label{def:RL1} \\
R_H^1 &
             \equiv & -\frac12(|\overline{z}_q|^2-\eta \Re\{
             z_q^*\overline{z}_qe^{i\phi_q} \})  = -\frac12 (R_H-A_D) \\
A_D^1 & \equiv & \frac12
             |\overline{z}_q|^2 = \frac12\left(\frac12(R_H+R_L)-A_D\right)  \\
\overline{R}_L^1 & \equiv &
             \frac12(|z_q|^2+\eta \Re\{ z_q^*\overline{z}_qe^{i\phi_q} \})
             = \frac12 (R_L+A_D)\\
\overline{R}_H^1 & \equiv & \frac12(|z_q|^2 - \eta \Re\{
             z_q^*\overline{z}_qe^{i\phi_q} \}) = \frac12 (R_H-A_D)\\
\overline{A}_D^1 &
             \equiv & \frac12 |z_q|^2 = \frac12\left(\frac12(R_H+R_L) A_D\right)
             \label{def:AM1bar}
\end{eqnarray}
With regard to the final state itself, a few comments are
necessary regarding the octet-singlet mixing in the $\eta$
system. This mixing, parametrized by the mixing angle $\theta_P$,
is still a controversial issue (for current experimental values, see e.g.\
\cite{europhys}), so
rather than using some specific value for $\theta_P$, we rewrite
the $\eta$ wavefunction in the following way: \label{sec:etameson}
\begin{eqnarray}
 \eta & = & \frac{1}{\sqrt{6}}(u\bar{u}+d\bar{d}-2s\bar{s})\cos(\theta_P) -
 \frac{1}{\sqrt{3}}(u\bar{u}+d\bar{d}+s\bar{s})\sin(\theta_P) \nonumber
 \\ & \equiv &
 N_\eta(u\bar{u}+d\bar{d}) + S_\eta (s\bar{s})
 \label{eq:etaparm}
\end{eqnarray}
\begin{equation}
N_\eta \equiv
\frac{\cos(\theta_P)}{\sqrt{6}}-\frac{\sin(\theta_P)}{\sqrt{3}}
\hspace*{1.5cm} S_\eta \equiv
\frac{-\sin(\theta_P)}{\sqrt{3}}-\frac{2\cos(\theta_P)}{\sqrt{6}}
\end{equation}
which is the definition we shall use in the following.
\subsection{The time-independent amplitudes}
The expressions given above for $|A_q(t)|^2$ and
$|\overline{A}_q(t)|^2$ depend on the time-independent amplitudes
parametrized by eqs.~(\ref{eq:Adef}) and (\ref{eq:Abardef}). In
non-leptonic processes, these amplitudes can generally not be
evaluated by any method relying on the factorization approach due
to final state interaction (FSI) effects. In the present work,
however, we do not need an explicit calculation. Rather, we wish
to obtain a parametrization of the $B_d$ and $B_s$ amplitudes
that will allow us to relate them by SU(3) flavour symmetry. To
arrive at such a parametrization, it is sufficient to use the
method of quark topologies which does not require the ability to
solve the full theory and which allows a systematic classification of
long-distance contributions (for a recent update on this method,
see \cite{non-leptonic}). With the topologies listed in
figure~\ref{tab:QT}, we obtain the following\footnote{The computational
  details can be found in an unpublished project. Please contact the author
  if a copy is needed.}:
\FIGURE[ht]{
\begin{fmffile}{QT1}
\setlength{\extrarowheight}{6pt}
\begin{tabular}{cccccc} & & & & & \\
\multicolumn{2}{c}{\hspace*{1cm}\begin{fmfgraph*}(55,45)
\fmfpen{thin}\fmfset{arrow_len}{2.5mm}
        \fmfset{curly_len}{2mm}\fmfset{wiggly_len}{2mm}
\fmfforce{0,.43h}{B_low} \fmfforce{0,.58h}{B_up} %
\fmfforce{.7w,0}{eta_low} \fmfforce{0.79w,0.09h}{eta_up}
\fmf{phantom,label=$\eta$,label.side=right}{eta_low,eta_up}
\fmfforce{.5w,.48h}{v1} 
\fmfforce{.50w,.49h}{circlecenter}
\fmfforce{.52w,.62h}{ccbararcl}
\fmfforce{.61w,.53h}{ccbararcr}
\fmfforce{.83w,.99h}{ccbartopl}
\fmfforce{.9w,.88h}{ccbartopr}
\fmf{phantom,label=$J/\psi$,label.side=left}{ccbartopl,ccbartopr}
\fmfforce{0,.5h}{labelBb}
\fmf{fermion,tension=1,right=.15}{eta_up,vbd}
\fmf{fermion,tension=1,right=.15}{vbd,B_up}
\fmf{fermion,left=.15,tension=1}{B_low,vdd,eta_low}
\fmf{phantom,tension=10}{v1,vbd}
\fmf{phantom,tension=2}{v1,vdd} 
\fmfv{label=$B_{d/s}$,label.dist=2}{labelBb}
\fmfv{decor.shape=circle,decor.filled=30,decor.size=.11w}{circlecenter}
\fmf{plain,tension=1.5,right}{ccbararcl,ccbararcr}
\fmf{fermion}{ccbartopl,ccbararcl}
\fmf{fermion}{ccbararcr,ccbartopr}
\end{fmfgraph*}
\hspace*{1cm}}  &
\multicolumn{2}{c}{\begin{fmfgraph*}(55,45)
\fmfpen{thin}\fmfset{arrow_len}{2.5mm}
        \fmfset{curly_len}{2mm}\fmfset{wiggly_len}{2mm}
\fmfforce{0,.63h}{B_low} \fmfforce{0,.77h}{B_up} %
\fmfforce{.85w,0}{eta_low} \fmfforce{0.94w,0.09h}{eta_up}
\fmf{phantom,label=$\eta$,label.side=left,label.dist=2}{eta_up,eta_low}
\fmfforce{.45w,.7h}{circlecenter}
\fmfforce{.36w,.63h}{v_an_low}
\fmfforce{.36w,.77h}{v_an_high}
\fmfforce{.54w,.63h}{v_em_low}
\fmfforce{.54w,.77h}{v_em_high}
\fmfforce{.95w,.77h}{ccbar_high}
\fmfforce{.95w,.63h}{ccbar_low}
\fmf{phantom,label=$J/\psi$,label.side=right,label.dist=2}{ccbar_low,ccbar_high}
\fmfforce{0,.7h}{labelBb}
\fmf{fermion}{B_low,v_an_low}
\fmf{plain,left=.3}{v_an_low,v_em_low}
\fmf{fermion}{v_an_high,B_up}
\fmfv{label=$B_{d/s}$,label.dist=2}{labelBb}
\fmfv{decor.shape=circle,decor.filled=30,decor.size=.11w}{circlecenter}
\fmf{fermion}{ccbar_high,v_em_high}
\fmf{plain,left=0.3}{v_em_high,v_an_high}
\fmf{fermion}{v_em_low,ccbar_low}
\fmfforce{.68w,.5h}{v_eta_close}
\fmfforce{.59w,.39h}{v_eta_far}
\fmf{fermion}{eta_up,v_eta_close}
\fmf{plain,right,tension=1}{v_eta_close,v_eta_far}
\fmf{fermion}{v_eta_far,eta_low}

\end{fmfgraph*}
} & \multicolumn{2}{c}{\begin{fmfgraph*}(55,45)
\fmfpen{thin}\fmfset{arrow_len}{2.5mm}
        \fmfset{curly_len}{2mm}\fmfset{wiggly_len}{2mm}
\fmfforce{0,.63h}{B_low} \fmfforce{0,.77h}{B_up} %
\fmfforce{.85w,0}{eta_low} \fmfforce{0.94w,0.09h}{eta_up}
\fmf{phantom,label=$J/\psi$,label.side=left,label.dist=2}{eta_up,eta_low}
\fmfforce{.45w,.7h}{circlecenter}
\fmfforce{.36w,.63h}{v_an_low}
\fmfforce{.36w,.77h}{v_an_high}
\fmfforce{.54w,.63h}{v_em_low}
\fmfforce{.54w,.77h}{v_em_high}
\fmfforce{.95w,.77h}{ccbar_high}
\fmfforce{.95w,.63h}{ccbar_low}
\fmf{phantom,label=$\eta$,label.side=right,label.dist=2}{ccbar_low,ccbar_high}
\fmfforce{0,.7h}{labelBb}
\fmf{fermion}{B_low,v_an_low}
\fmf{plain,left=0.3}{v_an_low,v_em_low}
\fmf{fermion}{v_an_high,B_up}
\fmfv{label=$B_{d/s}$,label.dist=2}{labelBb}
\fmfv{decor.shape=circle,decor.filled=30,decor.size=.11w}{circlecenter}
\fmf{fermion}{ccbar_high,v_em_high}
\fmf{plain,left=.3}{v_em_high,v_an_high}
\fmf{fermion}{v_em_low,ccbar_low}
\fmfforce{.68w,.5h}{v_eta_close}
\fmfforce{.59w,.39h}{v_eta_far}
\fmf{fermion}{eta_up,v_eta_close}
\fmf{plain,right,tension=1}{v_eta_close,v_eta_far}
\fmf{fermion}{v_eta_far,eta_low}
\end{fmfgraph*}
} \\
\multicolumn{2}{c}{\textsf{EMISSION}} & \multicolumn{2}{c}{\textsf{EMISSION
  ANNIHILATION 1}} & \multicolumn{2}{c}{\textsf{EMISSION ANNIHILATION 2}} \\
\hspace*{1.5cm} & \hspace*{1.5cm} & \hspace*{1.5cm} & \hspace*{1.5cm} &
\hspace*{1.5cm} & \hspace*{1.5cm}\\
 & \multicolumn{2}{c}{\begin{fmfgraph*}(55,45)
\fmfpen{thin}\fmfset{arrow_len}{2.5mm}
        \fmfset{curly_len}{2mm}\fmfset{wiggly_len}{2mm}
\fmfforce{0,.43h}{B_low} \fmfforce{0,.58h}{B_up} %
\fmfforce{.7w,0}{eta_low} \fmfforce{0.79w,0.09h}{eta_up}
\fmf{phantom,label=$\eta$,label.side=right}{eta_low,eta_up}
\fmfforce{.5w,.48h}{v1} 
\fmfforce{.45w,.42h}{circlecenter}
\fmfforce{.48w,.61h}{ccbararcl}
\fmfforce{.57w,.53h}{ccbararcr}
\fmfforce{.71w,.99h}{ccbartopl}
\fmfforce{.78w,.88h}{ccbartopr}
\fmfforce{.24w,.1h}{vdd}
\fmfforce{.44w,.37h}{nothercircleT}
\fmfforce{.33w,.2h}{nothercircleB}

\fmf{phantom,label=$J/\psi$,label.side=left}{ccbartopl,ccbartopr}
\fmfforce{0,.5h}{labelBb}
\fmf{fermion,tension=1,right=.15}{eta_up,vbd}
\fmf{fermion,tension=1,right=.15}{vbd,B_up}
\fmf{fermion,right=.15,tension=4}{B_low,vdd,eta_low}
\fmf{phantom,tension=10}{v1,vbd}

\fmf{fermion,tension=0.5,left}{nothercircleT,nothercircleB,nothercircleT}

\fmfv{label=$B_{d/s}$,label.dist=3}{labelBb}
\fmfv{decor.shape=circle,decor.filled=30,decor.size=.11w}{circlecenter}
\fmf{plain,tension=1.5,right}{ccbararcl,ccbararcr}
\fmf{fermion}{ccbartopl,ccbararcl}
\fmf{fermion}{ccbararcr,ccbartopr}
\end{fmfgraph*}
} & \multicolumn{2}{c}{\begin{fmfgraph*}(55,45)
\fmfpen{thin}\fmfset{arrow_len}{2.5mm}
        \fmfset{curly_len}{2mm}\fmfset{wiggly_len}{2mm}
\fmfforce{0,.43h}{B_low} \fmfforce{0,.58h}{B_up} %
\fmfforce{.45w,.5h}{v1} 
\fmfforce{.51w,.5h}{circlecenter}
\fmfforce{.48w,.5h}{nothercircleR}
\fmfforce{.34w,.5h}{nothercircleL}
\fmfforce{.65w,.7h}{ccbar_arcl}
\fmfforce{.74w,.63h}{ccbar_arcr}
\fmfforce{.9w,.95h}{ccbar_topl}
\fmfforce{.99w,.88h}{ccbar_topr}
\fmfforce{.65w,.3h}{eta_arcl}
\fmfforce{.74w,.37h}{eta_arcr}
\fmfforce{.9w,.05h}{eta_botl}
\fmfforce{.99w,.12h}{eta_botr}
\fmf{phantom,label=$J/\psi$,label.side=left}{ccbar_topl,ccbar_topr}
\fmf{phantom,label=$\eta$,label.side=right}{eta_botl,eta_botr}
\fmfforce{0,.5h}{labelBb}
\fmfforce{.3w,.65h}{B_up1}
\fmfforce{.3w,.35h}{B_low1}
\fmfforce{.55w,.5h}{vbd}
\fmf{fermion,tension=1,left=.25}{B_low,B_low1}
\fmf{plain,tension=1,right=.6}{B_low1,vbd}

\fmf{fermion,tension=1,left=.25}{B_up1,B_up}
\fmf{plain,tension=1,right=.6}{vbd,B_up1}
\fmf{phantom,tension=8}{v1,vbd}
\fmfv{label=$B_{d/s}$,label.dist=2}{labelBb}
\fmfv{decor.shape=circle,decor.filled=30,decor.size=.11w}{circlecenter}
\fmf{fermion,tension=0.5,left}{nothercircleL,nothercircleR,nothercircleL}
\fmf{fermion}{ccbar_topl,ccbar_arcl}
\fmf{fermion}{ccbar_arcr,ccbar_topr}
\fmf{fermion,tension=1.5,right}{ccbar_arcl,ccbar_arcr}
\fmf{fermion}{eta_botl,eta_arcl}
\fmf{fermion}{eta_arcr,eta_botr}
\fmf{fermion,tension=1.5,left}{eta_arcl,eta_arcr}
\end{fmfgraph*}
} & \\
& \multicolumn{2}{c}{\textsf{PENGUIN EMISSION}} &
\multicolumn{2}{c}{\textsf{DOUBLE PENGUIN ANNIHILATION}} &
\end{tabular}\end{fmffile}\vspace*{-3mm}
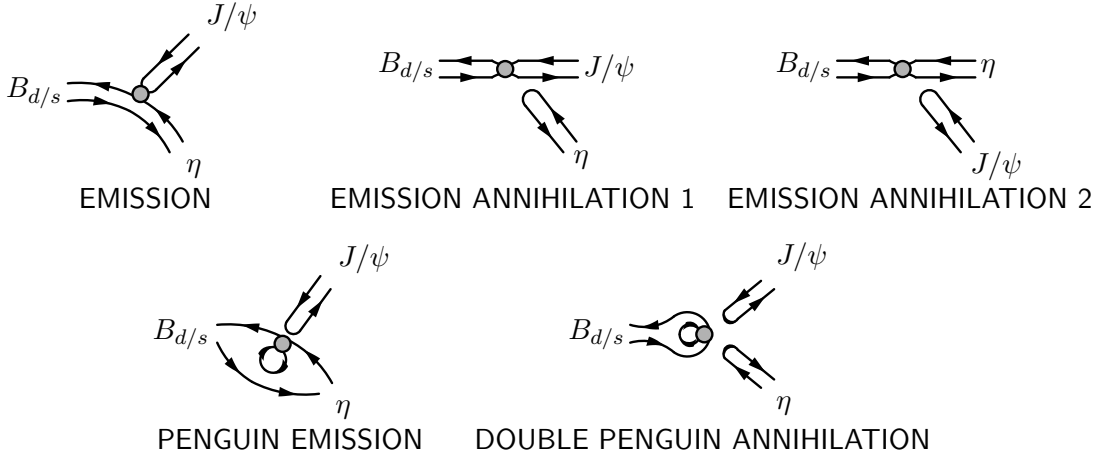
\caption{Quark topologies in \BdsJpsi. The gray blobs denote the contracted
  short-distance parts.\label{tab:QT}}}
\begin{eqnarray}
A(\BdJpsi) & = & N_\eta\left(\lbd{c} A_d + \lbd{u} B_d\right) +
\lbd{u}N_\eta\left[D_d + \frac{\lbd{c}}{\lbd{u}}\zeta_d + \xi_d
\right]\label{eq:simpampd}\\
A(\BsJpsi) & = & S_\eta\left(\lbs{c} A_s + \lbs{u} B_s\right) +
\lbs{u}N_\eta\left[D_s + \frac{\lbs{c}}{\lbs{u}}\zeta_s + \xi_s
\right] \label{eq:simpamps}
\end{eqnarray}
with the definitions:
\begin{eqnarray}
A_x & \equiv & \expt{1,2}{ccx}{E} + \expt{1,2}{ccx}{PE} +
\expen{x}{c}{E} +
\smsum{q}\expen{x}{q}{PE}\\
B_x & \equiv & \expt{1,2}{uux}{PE} + \expen{x}{c}{E} + \smsum{q}
\expen{x}{q}{PE}\\
D_x & \equiv & \expt{1,2}{uux}{EA2}\\
\zeta_x & \equiv &
\frac{2N_\eta+S_\eta}{N_\eta}\left(\expt{1,2}{ccx}{EA1} +
\expt{1,2}{ccx}{DPA} + \expen{x}{c}{EA1}
+ \smsum{q}\expen{x}{q}{DPA} \right) \nonumber\\
& & \hspace{15mm}+ \expen{x}{u}{EA2}+ \expen{x}{d}{EA2}+
\scriptstyle{\frac{S_\eta}{N_\eta}}\displaystyle\expen{x}{s}{EA2}
      \label{eq:genamppar1}\\
\xi_x & \equiv &
\frac{2N_\eta+S_\eta}{N_\eta}\left(\expt{1,2}{uux}{DPA} +
\expen{x}{c}{EA1} + \smsum{q}\expen{x}{q}{DPA} \right) \nonumber\\
& & \hspace{15mm}+\expen{x}{u}{EA2}+\expen{x}{d}{EA2}+
\scriptstyle{\frac{S_\eta}{N_\eta}}\displaystyle\expen{x}{s}{EA2}
\label{eq:genamppar2}
\end{eqnarray}
where $\lambda_{bq}^{q'} \equiv V_{q'b}^*V_{q'q}$,
$\expt{i}{uud}{X}$ denotes the insertion of the OPE operator
\OPE{i} having external quark lines $buud$ into topology
\topl{X}, and \expt{pen}{(u)d}{X} denotes the combined
contribution of the QCD penguin operators \OPE{3-6}. The
quark content is denoted $(q)d,(q)s$ where $q$ is the flavour of
the $q\bar{q}$ pair coming from the gluon and $d$ or $s$ are from the
flavour-changing $b$-quark transition. It should be mentioned
that, relative to the current-current operators \OPE{1,2}, the
electroweak (QCD) penguins contain an extra power of $\alpha_{EM}$
($\alpha_s$). In neither the $B_d$ nor the $B_s$ case are the dominant
current-current contributions CKM suppressed relative to the penguins, and so
we expect $|A_{EWP}|/|A_{CC}|=\mathcal{O}(10^{-2})=|A_{EWP}|/|A_{QCDP}|$. I
have therefore neglected elelectroweak penguin contributions in the above
analysis.  

Noting that the $\zeta$ and $\xi$ terms are SU(3) suppressed and
that the $D$ terms are OZI suppressed (see figure~\ref{tab:QT}),
we neglect these and obtain:
\begin{eqnarray}
\mathcal{A}_d = A(\BdJpsi) & = &
 N_\eta\left( V_{cb}^*V_{cd}A_d + V_{ub}^*V_{ud}B_d \right) \nonumber \\
 & \equiv & N_d \left[ 1-{a}_de^{i{\theta}_d}e^{i\gamma}\right] \label{eq:finalparmd}\\
\mathcal{A}_s = A(\BsJpsi) & = &
 S_\eta\left( V_{cb}^*V_{cs}A_s + V_{ub}^*V_{us}B_s \right) \nonumber \\
& \equiv &
N_s\left[1-{a}_s{\textstyle{\frac{\lambda^2}{1-\lambda^2}}}e^{i{\theta}_s}e^{i\gamma}\right]
 \label{eq:finalparms}
\end{eqnarray}
with
\begin{eqnarray}
N_d \equiv -N_\eta A \lambda^3 A_d\hspace{3mm} , & \hspace{3mm}
 {a}_d \equiv R_b
\left(1-\frac{\lambda^2}{2}\right)\left|\frac{B_d}{A_d}\right|\hspace{3mm}
, &\hspace{3mm} \theta_d \equiv
\mbox{Arg}\left\{\textstyle\frac{B_d}{A_d}\right\}
\label{eq:dparams}
\\
N_s \equiv S_\eta A \lambda^2 A_s
\left(1-\frac{\lambda^2}{2}\right)\hspace{3mm} , &\hspace{3mm}
{a}_s \equiv R_b
\left(1-\frac{\lambda^2}{2}\right)\left|\frac{B_s}{A_s}\right|
\hspace{3mm} , &\hspace{3mm} \theta_s \equiv -
\mbox{Arg}\left\{\textstyle\frac{B_s}{A_s}\right\}
\label{eq:sparams}
\end{eqnarray}
and the Wolfenstein parameters \cite{extractinggamma}:
\begin{eqnarray}
 \lambda \equiv |V_{us}|=0.22, \hspace{2.6mm} & \hspace{2.6mm}
A\equiv \frac{1}{\lambda^2}|V_{cb}|=0.81 \pm 0.06, \hspace{2.6mm}
& \hspace{2.6mm} R_b \equiv \frac{1}{\lambda} \left|
\frac{V_{ub}}{V_{cb}} \right| = 0.41 \pm 0.07
\end{eqnarray}
With eqs.~(\ref{eq:finalparmd}) and (\ref{eq:finalparms}) we have
recovered exactly the form of eq.~(\ref{eq:Adef}) by which we
parametrized the time-independent amplitudes, but in a form that
explicitly separates the CKM structure from the strong
amplitudes. This is of essential use below.


\section{Extracting $\gamma$ from \BdsJpsi}
\label{sec:extract} We here adapt the method proposed in
\cite{extractinggamma} for the decay considered here. Defining
the time-dependent $CP$ asymmetry by:
\begin{equation}
a_{CP} =
\frac{|A_q(t)|^2-|\overline{A}_q(t)|^2}{|A_q(t)|^2+|\overline{A}_q(t)|^2}
\end{equation}
and inserting the above expressions, one obtains
\cite{extractinggamma}:
\begin{equation}
a_{CP}(t) = 2 e^{-\Gamma t} \left[
  \frac{\mathcal{A}_{CP}^{dir} \cos (\DeltaMt\ ) + \mathcal{A}_{CP}^{mix}
  \sin (\DeltaMt\ )
    }{e^{-\Gamma_H t} + e^{-\Gamma_L t} + \mathcal{A}_{\Delta
    \Gamma } (e^{-\Gamma_H t} - e^{-\Gamma_L t} )
    } \right] \label{eq:aCP}
\end{equation}
with
\begin{eqnarray}
\mathcal{A}_{CP}^{dir} \equiv \frac{2A_D}{R_H + R_L} & = &
\frac{2 \tilde{a}_q \sin\gamma\sin\tilde{\theta}_q
}
  {1 - 2 \tilde{a}_q \cos\gamma\cos\tilde{\theta}_q
  +\tilde{a}_q^2} \label{def:Adir}\\
\mathcal{A}_{CP}^{mix} \equiv \frac{2A_M}{R_H + R_L}
& = & \eta \frac{\sin\phi -
2\tilde{a}_q\sin(\gamma+\phi)\cos\tilde{\theta}_q +
  \tilde{a}_q^2\sin(2\gamma+\phi)}{1 - 2 \tilde{a}_q \cos\gamma\cos\tilde{\theta}_q
  +\tilde{a}_q^2}
\label{def:Amix} \\
\mathcal{A}_{\Delta \Gamma} \equiv \frac{R_H - R_L}
{R_H + R_L}
& = & -\eta \frac{\cos\phi_q-2 \tilde{a}_q
\cos(\gamma+\phi_q)\cos\tilde{\theta}_q +
  \tilde{a}_q^2\cos(2\gamma +\phi_q)}{1 - 2 \tilde{a}_q \cos\gamma\cos\tilde{\theta}_q
  +\tilde{a}_q^2}
\label{def:ADeltaGamma}
\end{eqnarray}
\begin{eqnarray}
\tilde{a}_q & = & \left\{
\begin{array}[c]{cl}
a_d & \mbox{ \ \ ; for \BdJpsi} \\
a_s\frac{\lambda^2}{1-\lambda^2} & \mbox{ \ \ ; for \BsJpsi}
\end{array} \right. \\
\tilde{\theta}_q & = & \left\{
\begin{array}[c]{cl}
\theta_d & \mbox{ \ \ ; for \BdJpsi} \\
\theta_s + 180^\circ & \mbox{ \ \ ; for \BsJpsi}
\end{array} \right.
\end{eqnarray}
As $a_s$ is suppressed by $\frac{\lambda^2}{1-\lambda^2}$, we
expect a very small direct CP asymmetry in the $B_s$ process, thus
we shall use the asymmetries in \BdJpsi\ combined with $R_d$ and
$R_s$ for extracting $\gamma$ ($R_q\equiv\frac12 (R_H^q+R_L^q)$).
It should be mentioned that, due to the smallness of
$\Delta\Gamma_d$, the `observable' $\mathcal{A}_{\Delta\Gamma}$
will only be measurable for the $B_s$ system. This has no direct
consequence on the analysis presented here. Only two of the three
asymmetries are independent (see below), and so a measurement of
$\mathcal{A}^{dir}_{CP}$ and $\mathcal{A}^{mix}_{CP}$ is
sufficient.
\begin{equation}
(\mathcal{A}^{dir}_{CP})^2+(\mathcal{A}^{mix}_{CP})^2+(\mathcal{A}_{\Delta\Gamma})^2
= 1
\end{equation}
which has been checked also to be valid when going to $\epspq\ne 0$.

The observables $\mathcal{A}^{dir}_{CP}$ and 
$\mathcal{A}^{mix}_{CP}$ do not
depend on the normalization, $|N_d|^2$, and can in principle be
obtained by fitting to the $CP$ asymmetry. This yields two
equations (\ref{def:Adir})--(\ref{def:Amix}) in the three
``unknowns": $\tilde{a}_d$, $\tilde{\theta}_d$, and $\gamma$
(taking the mixing angle, $2\beta$, to be known beforehand). 
Thus, we need one more observable. Measuring the CP averaged
rate yields
\begin{equation}
\left< \Gamma_q \right> \equiv \Pi_2 \times |N_q|^2 \times R_q
\label{def:Gamma}
\end{equation}
where $\Pi_2$ is the 2-body phase space, and the normalization
factors for the $B_d$ and $B_s$ modes are given by
eqs.~(\ref{eq:dparams}) and (\ref{eq:sparams}). A priori, we
cannot determine this normalization, but assuming now that the
strong interaction dynamics is the same for the $B_d$ and $B_s$
modes ($U$-spin symmetry), we have $a_d=a_s\equiv a$,
$\theta_d=\theta_s\equiv\theta$, and taking kinematics into
account:
\begin{equation}
\left|\frac{N_s}{N_d}\right|^2 =
\frac{\lambda^2}{1-\lambda^2}\left|\frac{A_s}{A_d}\right|^2 =
\frac{\lambda^2}{1-\lambda^2}\frac{\lambda(m_{B_s}^2,m_{J/\psi}^2,m_\eta^2)}{
\lambda(m_{B_d}^2,m_{J/\psi}^2,m_\eta^2)}
\end{equation}
with the $\lambda$ function being the standard one used in
kinematics. Thus, forming the ratio $R_s/R_d=(R_H^s +
R_L^s)/(R_H^d + R_L^d)$, the strong dynamics cancels out, and we
find by combining eq.~(\ref{def:Gamma}) with $R = \frac12 (R_H +
R_L) = \frac12 (1 - 2 \tilde{a}_q \cos\gamma\cos\tilde{\theta}_q
+\tilde{a}_q^2)$:
\begin{eqnarray}
\frac{R_s}{R_d} \equiv H & = & \frac{1 -
2{\textstyle\frac{\lambda^2}{1-\lambda^2}a_s\cos\gamma\cos\theta_s
    +(\frac{\lambda^2}{1-\lambda^2})^2a_s^2 }}
     {1-2a_d\cos\gamma\cos\theta_d + a_d^2} \nonumber \\
& = & \frac{\left<\Gamma_s\right>}{\left<\Gamma_d\right>}
\frac{|N_d|^2}{|N_s|^2} \frac{m_{B_s}^3}{m_{B_d}^3}
\left(\frac{\lambda(m_{B_d}^2,m_{J/\psi}^2,m_\eta^2)}{
\lambda(m_{B_s}^2,m_{J/\psi}^2,m_\eta^2)} \right)^{1/2} \nonumber \\
& = & {\textstyle
       \frac{\lambda^2}{1-\lambda^2}}\frac{\left<\Gamma_s\right>}{\left<\Gamma_d\right>} \frac{|N_\eta|^2}{|S_\eta|^2}
       \frac{m_{B_s}^3}{m_{B_d}^3}\left(\frac{\lambda(m_{B_d}^2,m_{J/\psi}^2,m_\eta^2)}{
\lambda(m_{B_s}^2,m_{J/\psi}^2,m_\eta^2)} \right)^{3/2}
\label{eq:rsdivrd}
\end{eqnarray}
This relation furnishes us with the last observable we need for
determining $a$, $\theta$, and $\gamma$ as functions of the
mixing phase, $\phi$. In the Standard Model, this angle is
negligibly small for the $B_s - \overline{B}_s$ system whereas it
is $2\beta$ for the $B_d - \overline{B}_d$ system. Thus, the
extraction of $\gamma$ requires $\beta$ as an input parameter.

To summarize: The two observables in eqs.\
(\ref{def:Adir}) and (\ref{def:Amix}) are to be determined
for the decay \BdJpsi\ by measuring the CP asymmetry and fitting
to the time-dependent decay amplitudes. This will require tagging
and will fix a contour in the $\gamma - a$ plane. The last
observable is provided by $H\equiv R_s/R_d$ which does not
require tagging as it depends only on the CP averaged rates.
Together with e.g.\ $\mathcal{A}^{mix}_{CP}$, we fix another
contour in the $\gamma - a$ plane. The intersections of these two
contours will fix both $a$ and $\gamma$, to a theoretical
precision depending on SU(3)-breaking corrections and the
accuracies on the $\eta$ and $B_d$ mixing angles, $\theta_P$ and
$2\beta$.

As we have chosen the same terminology and parametrizations as
\cite{extractinggamma} the contour equations in that paper are
directly applicable, and we do not repeat them here.


\section{Branchings for \BdsJpsi}
\label{sec:brbjpsi} We now present a simplified picture of the
\BdsJpsi\ transitions in order to obtain an estimate for the
branching ratios. We again take the strong interaction dynamics
symmetric under SU(3) transformations and further make the simplifying ansatz
of fig.~\ref{fig:simpsu3} that the $\OPE{1,2}$ insertions into emission
topologies represent the dominant contributions, resulting in the parameters
$a_d$ and $a_s$ of the previous sections being negligibly small (we really only
have to make 
this crude assumption for the \BdJpsi\ and $\BsJpsiK_S$ processes. In
$\BsJpsi$ and $\BdJpsiK_S$ (see \cite{extractinggamma}), we can
justify neglecting these terms because of the
$\lambda^2$ suppression). 
With these assumptions, the amplitudes in fig.~\ref{fig:simpsu3} can differ
only by CKM factors, kinematics and factors coming from the hadronic
wavefunctions. 
\FIGURE[ht]{ \setlength{\extrarowheight}{0pt}
\begin{tabular}{cc}
\begin{fmffile}{su3_1} \begin{fmfgraph*}(90,90)
\fmfpen{thin}\fmfset{arrow_len}{3mm}
\fmfset{curly_len}{2mm}\fmfset{wiggly_len}{2mm}
\fmfforce{0,.2h}{Bslow}\fmfforce{w,.2h}{etalow}
\fmfforce{0,.3h}{Bsmid}\fmfforce{w,.3h}{etamid}
\fmfforce{0,.4h}{Bshigh}\fmfforce{w,.4h}{etahigh}
\fmfforce{.93w,.83h}{jpsimid}
\fmfforce{.86w,.9h}{jpsileft}
\fmfforce{w,.76h}{jpsiright}
\fmf{fermion,label=$\scriptstyle s$,label.side=right,label.dist=3}{Bslow,etalow}
\fmf{fermion,label=$\scriptstyle \bar{s}$,label.dist=3}{etahigh,v1}
\fmf{fermion,label=$\scriptstyle \bar{b}$,label.dist=3}{v1,Bshigh}
\fmffreeze
\fmf{fermion,label=$\scriptstyle c$,label.side=left,label.dist=3}{v1,jpsileft}
\fmf{fermion,label=$\scriptstyle \bar{c}$,label.side=left,label.dist=3}{jpsiright,v1}
\fmfv{label=$\scriptstyle B_s$,label.dist=0}{Bsmid}
\fmfv{label=$\scriptstyle \eta$,label.dist=0}{etamid}
\fmfv{label=$\scriptstyle J\! /\! \psi$,label.dist=0}{jpsimid}
\fmf{plain,left=.55}{Bslow,Bshigh}
\fmf{plain,left=.55}{Bshigh,Bslow}
\fmf{plain,left=.55}{etalow,etahigh}
\fmf{plain,left=.55}{etahigh,etalow}
\fmf{plain,left=.55}{jpsileft,jpsiright}
\fmf{plain,left=.55}{jpsiright,jpsileft}
\end{fmfgraph*}

 \end{fmffile} \hspace{3mm}
&\hspace{3mm} \begin{fmffile}{su3_2} \begin{fmfgraph*}(90,90)
\fmfpen{thin}\fmfset{arrow_len}{3mm}
\fmfset{curly_len}{2mm}\fmfset{wiggly_len}{2mm}
\fmfforce{0,.2h}{Bslow}\fmfforce{w,.2h}{etalow}
\fmfforce{0,.3h}{Bsmid}\fmfforce{w,.3h}{etamid}
\fmfforce{0,.4h}{Bshigh}\fmfforce{w,.4h}{etahigh}
\fmfforce{.93w,.83h}{jpsimid}
\fmfforce{.86w,.9h}{jpsileft}
\fmfforce{w,.76h}{jpsiright}
\fmf{fermion,label=$\scriptstyle d$,label.side=right,label.dist=3}{Bslow,etalow}
\fmf{fermion,label=$\scriptstyle \bar{s}$,label.dist=3}{etahigh,v1}
\fmf{fermion,label=$\scriptstyle \bar{b}$,label.dist=3}{v1,Bshigh}
\fmffreeze
\fmf{fermion,label=$\scriptstyle c$,label.side=left,label.dist=3}{v1,jpsileft}
\fmf{fermion,label=$\scriptstyle \bar{c}$,label.side=left,label.dist=3}{jpsiright,v1}
\fmfv{label=$\scriptstyle B_d$,label.dist=0}{Bsmid}
\fmfv{label=$\scriptstyle K^0$,label.dist=0}{etamid}
\fmfv{label=$\scriptstyle J\! /\! \psi$,label.dist=0}{jpsimid}
\fmf{plain,left=.55}{Bslow,Bshigh}
\fmf{plain,left=.55}{Bshigh,Bslow}
\fmf{plain,left=.55}{etalow,etahigh}
\fmf{plain,left=.55}{etahigh,etalow}
\fmf{plain,left=.55}{jpsileft,jpsiright}
\fmf{plain,left=.55}{jpsiright,jpsileft}
\end{fmfgraph*}

 \end{fmffile} \\
\begin{fmffile}{su3_3} \begin{fmfgraph*}(90,90)
\fmfpen{thin}\fmfset{arrow_len}{3mm}
\fmfset{curly_len}{2mm}\fmfset{wiggly_len}{2mm}
\fmfforce{0,.2h}{Bslow}\fmfforce{w,.2h}{etalow}
\fmfforce{0,.3h}{Bsmid}\fmfforce{w,.3h}{etamid}
\fmfforce{0,.4h}{Bshigh}\fmfforce{w,.4h}{etahigh}
\fmfforce{.93w,.83h}{jpsimid}
\fmfforce{.86w,.9h}{jpsileft}
\fmfforce{w,.76h}{jpsiright}
\fmf{fermion,label=$\scriptstyle d$,label.side=right,label.dist=3}{Bslow,etalow}
\fmf{fermion,label=$\scriptstyle \bar{d}$,label.dist=3}{etahigh,v1}
\fmf{fermion,label=$\scriptstyle \bar{b}$,label.dist=3}{v1,Bshigh}
\fmffreeze
\fmf{fermion,label=$\scriptstyle c$,label.side=left,label.dist=3}{v1,jpsileft}
\fmf{fermion,label=$\scriptstyle \bar{c}$,label.side=left,label.dist=3}{jpsiright,v1}
\fmfv{label=$\scriptstyle B_d$,label.dist=0}{Bsmid}
\fmfv{label=$\scriptstyle \eta$,label.dist=0}{etamid}
\fmfv{label=$\scriptstyle J\! /\! \psi$,label.dist=0}{jpsimid}
\fmf{plain,left=.55}{Bslow,Bshigh}
\fmf{plain,left=.55}{Bshigh,Bslow}
\fmf{plain,left=.55}{etalow,etahigh}
\fmf{plain,left=.55}{etahigh,etalow}
\fmf{plain,left=.55}{jpsileft,jpsiright}
\fmf{plain,left=.55}{jpsiright,jpsileft}
\end{fmfgraph*}

 \end{fmffile} \hspace{3mm}
& \hspace{3mm} \begin{fmffile}{su3_4} \begin{fmfgraph*}(90,90)
\fmfpen{thin}\fmfset{arrow_len}{3mm}
\fmfset{curly_len}{2mm}\fmfset{wiggly_len}{2mm}
\fmfforce{0,.2h}{Bslow}\fmfforce{w,.2h}{etalow}
\fmfforce{0,.3h}{Bsmid}\fmfforce{w,.3h}{etamid}
\fmfforce{0,.4h}{Bshigh}\fmfforce{w,.4h}{etahigh}
\fmfforce{.93w,.83h}{jpsimid}
\fmfforce{.86w,.9h}{jpsileft}
\fmfforce{w,.76h}{jpsiright}
\fmf{fermion,label=$\scriptstyle s$,label.side=right,label.dist=3}{Bslow,etalow}
\fmf{fermion,label=$\scriptstyle \bar{d}$,label.dist=3}{etahigh,v1}
\fmf{fermion,label=$\scriptstyle \bar{b}$,label.dist=3}{v1,Bshigh}
\fmffreeze
\fmf{fermion,label=$\scriptstyle c$,label.side=left,label.dist=3}{v1,jpsileft}
\fmf{fermion,label=$\scriptstyle \bar{c}$,label.side=left,label.dist=3}{jpsiright,v1}
\fmfv{label=$\scriptstyle B_s$,label.dist=0}{Bsmid}
\fmfv{label=$\scriptstyle \overline{K}^0$,label.dist=0}{etamid}
\fmfv{label=$\scriptstyle J\! /\! \psi$,label.dist=0}{jpsimid}
\fmf{plain,left=.55}{Bslow,Bshigh}
\fmf{plain,left=.55}{Bshigh,Bslow}
\fmf{plain,left=.55}{etalow,etahigh}
\fmf{plain,left=.55}{etahigh,etalow}
\fmf{plain,left=.55}{jpsileft,jpsiright}
\fmf{plain,left=.55}{jpsiright,jpsileft}
\end{fmfgraph*}

 \end{fmffile}\\
\end{tabular}\vspace*{-3mm}
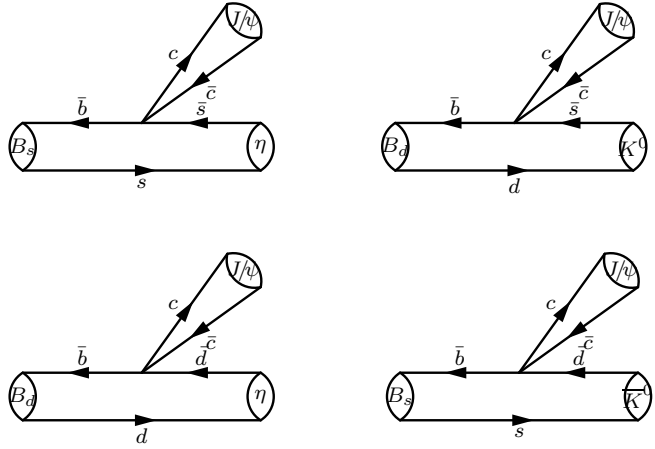
\caption{A simplified picture of the four SU(3) related
decays \BdsJpsi\ and
  $B_{d,s}\to J/\psi\ \! K^0 $. \label{fig:simpsu3}}
}
Indeed, eqs.~(\ref{eq:finalparmd}) and (\ref{eq:finalparms}) become
(separating
kinematics and dynamics):
\begin{eqnarray}
A(\BdJpsi) & = & N_\eta \lambda^c_{bd} A_d = N_\eta\lambda^c_{bd}
(p_{B_d}+p_\eta)^\mu\epsilon_\mu F(\BdJpsi) \\ 
A(\BsJpsi) & = & S_\eta\lambda^c_{bs}A_s = S_\eta \lambda^c_{bs}
(p_{B_s}+p_\eta)^\mu\epsilon_\mu F(\BsJpsi)
\end{eqnarray}
where $p$ and $\epsilon$ are momenta and polarization vectors, respectively,
and the form factors, $F$, parametrize the strong dynamics.

For the $\BdsJpsiK_S$ I use the amplitudes in \cite{extractinggamma} with the
same assumptions as above and with a slight modification of the author's
notation.
\begin{eqnarray}
A(\BdJpsiK_S) & = & \lambda^c_{bs} A_d =
\lambda^c_{bs}(p_{B_d}+p_{K_S})^\mu\epsilon_\mu F(\BdJpsiK_S) \\
A(\BsJpsiK_S) & = & \lambda^c_{bd} A_s = 
\lambda^c_{bd}(p_{B_s}+p_{K_S})^\mu\epsilon_\mu F(\BsJpsiK_S) 
\end{eqnarray}
With the assumption that the strong dynamics is SU(3) symmetric, the form
factors cancel out when forming ratios, and we arrive at:
\begin{eqnarray}
\frac{A(\BdJpsi)}{A(\BdJpsiK_S)} & = & 
N_\eta \sqrt{2} \frac{\lambda_{bd}^c}{\lambda_{bs}^c} \frac
{(p_{B_d}+p_\eta)^\mu\epsilon_\mu}
{(p_{B_d}+p_{K_S})^\mu\epsilon_\mu}\\
\frac{A(\BsJpsi)}{A(\BdJpsiK_S)} & = & 
S_\eta \sqrt{2} \frac
{(p_{B_s}+p_\eta)^\mu\epsilon_\mu}{(p_{B_d}+p_{K_S})^\mu\epsilon_\mu}
\\ 
\frac{A(\BsJpsiK_S)}{A(\BdJpsiK_S)} & = &
\frac{\lambda_{bd}^c}{\lambda_{bs}^c}\frac
{(p_{B_s}+p_{K_S})^\mu\epsilon_\mu}
{(p_{B_d}+p_{K_S})^\mu\epsilon_\mu}
\end{eqnarray}
where the $\sqrt{2}$ comes from the translation from $K^0$ to $K_S$. 
In addition to the modes we are interested in, we have written down
$A(\BsJpsiK_S)$ as a bonus. 

By using these relations, it is straightforward to obtain the
branching ratios for \BdsJpsi\ and $\BsJpsiK_S$ in terms of that
for $\BdJpsiK_S$ with the measured value $BR(\BdJpsiK^0 ) =
2BR(\BdJpsiK_S ) = (8.9\pm 1.2)\times 10^{-4} $ \cite{europhys}.
Inserting this value yields:
\begin{eqnarray}
BR_{\BdJpsi} & = & \textstyle{9\times 10^{-4}|N_\eta|^2
  \left|\frac{V_{cd}}{V_{cs}}\right|^2 \! \left(
  \frac{\lambda(m^2_{B_d}, m_{J/\psi}^2, m_{\eta}^2)}{\lambda(m^2_{B_d},
  m_{J/\psi}^2, m_K^2)} \right)^{3/2}} \\
BR_{\BsJpsi} & = & \textstyle{9\times 10^{-4} |S_\eta|^2
  \frac{m^3_{B_d}}{m^3_{B_s}}\!\left(
  \frac{\lambda(m^2_{B_s}, m_{J/\psi}^2, m_{\eta}^2)}{\lambda(m^2_{B_d},
  m_{J/\psi}^2, m_K^2)} \right)^{3/2}} \\
BR_{\BsJpsiK^0} & = & \textstyle{9\times 10^{-4}
  \left|\frac{V_{cd}}{V_{cs}}\right|^2 \frac{m^3_{B_d}}{m^3_{B_s}} \!\left(
  \frac{\lambda(m^2_{B_s}, m_{J/\psi}^2, m_{K}^2)}{\lambda(m^2_{B_d},
  m_{J/\psi}^2, m_K^2)} \right)^{3/2}}
\end{eqnarray}
As mentioned in section \ref{sec:etameson}, there is some
controversy as to the precise value of the $\eta$ mixing angle,
$\theta_P$, which determines $N_\eta$ and $S_\eta$. Rather than
adopting some specific value, we have varied the parameter between
 $-20^\circ < \theta_P < -10^\circ$, producing the results shown in table
\ref{tab:BRBJpsi}. The $\theta_P$-dependence between these limits is 
 linear to a good approximation. The uncertainty on these branching ratios
 is roughly 40\%, slightly more for $\BdJpsi$ and $\BsJpsiK_S$.
\TABLE[htb]{
\vspace*{0mm}\ \\
\begin{tabular}{|l|r|r|} \hline
    & \small{$\theta_P = -10^\circ$} & \small{$\theta_P = -20^\circ$}
    \\ \hline  $BR(B_d \to J/\psi\!\ \eta)$ &\small{$1.1\times 10^{-5}$}
&\small{$1.5\times 10^{-5} $}
    \\ $BR(B_s \to J/\psi\!\ \eta)$ &\small{$5.0\times 10^{-4}$}
&\small{$3.3\times 10^{-4}$}
    \\ \hline $BR(B_s\to J/\psi \overline{K}^0)$ &
    \multicolumn{2}{|c|}{$5.28\times 10^{-5}$} \\ \hline\end{tabular}
\caption{Estimated branching ratios for $\BdsJpsi$ and
$\BsJpsiKbar$. \label{tab:BRBJpsi}}}

According to the PYTHIA simulation for HERA-B
\cite{HERA-B_Proposal}, the production rate of $B_d$ mesons is
about five times greater than that of $B_s$ mesons, and so we
will expect to see about 5 times more $B_s$ decays than $B_d$
decays in the experiment for $\theta_P=-20^\circ$. For
$\theta_P=-10^\circ$, we expect about 10 times more $B_s$ decays
than $B_d$ decays. This situation is slightly better than for the
$\BdsJpsiK_S$ decays \cite{extractinggamma} where we expect to
see around 250 $B_d$ events for each $B_s$ event, and so the
statistical error on the $B_s$ events is going to have a larger
influence on the precision with which $\gamma$ can be extracted. In both
strategies, the asymmetries are to be determined for the mode with
\emph{least} statistics, so a few more factors are definitely of use.
\TABLE[ht]{
    \begin{tabular}{|lr|r|r|}\hline
    & & \small{$B_{d} \to J/\psi\!\ \eta$} & \small{$B_{s} \to
    J/\psi\!\ \eta$}
    \\ \hline  HERA-B &
                 \small{$\theta_P=-10^\circ$} &\small{100}
               & \small{850} \\(untagged, /yr) &
                 \small{$\theta_P=-20^\circ$} &\small{125}
               & \small{560} \\ \hline
               CDF II & \small{$\theta_P=-10^\circ$} &\small{700}
               & \small{$5.8\times 10^3$} \\
               (untagged, 2$\mbox{fb}^{-1}$) &
                 \small{$\theta_P=-20^\circ$} &\small{1000}
               & \small{$3.8\times 10^3$} \\ \hline
               ATLAS &
                 \small{$\theta_P=-10^\circ$}
               & \small{$4\times 10^4$}
               & \small{$3.5\times 10^5$} \\
               (tagged, 30$\mbox{fb}^{-1}$) & \small{$\theta_P=-20^\circ$}
               &\small{$5\times 10^4$}
    &\small{$2.5\times 10^5$} \\ \hline
     \multicolumn{4}{|c|}{\small $\eta$ not reconstructed ($\to$ factor
    $10\%-20\%$)}\\ \hline
    \end{tabular}
\caption{Estimated number of reconstructed \BdsJpsi\ events at HERA-B, CDFII, and
ATLAS.\label{tab:BJpsirec}}}
Given the branching ratios, it has been estimated how many events
will be seen by HERA-B, CDF-II, and ATLAS, based on their
simulations for $\BdJpsiK_S$
\cite{HERA-B_design}\cite{CDFII_TDR}\cite{ATLAS_TDR}. Results are listed in
table \ref{tab:BJpsirec}. As the
$\eta$ reconstruction efficiency is not known at present, the
numbers presented here are without $\eta$ reconstruction
included. A loose estimate of this efficiency is 10--20\%.

\noindent Taking the $\eta$ reconstruction efficiency into account, it is certain that
HERA-B will not be able to access this mode (the number given is for the
machine running at full luminosity), it is an open question whether CDF-II
will be able to get it (depending on how much luminosity they get before LHC,
and whether they improve their trigger efficiency
\cite{CDFII_TDR}\cite{CDFII_upgrade}, and it is certain that ATLAS will
access it within the first three years of operation.

\section{Branchings for \Bdsetallp}
\label{sec:bretallp} The branchings for \Bsetallp\ can be obtained
by using that the process is related by the approximate SU(3) flavour
symmetry to $\BdKllp$ whose spectrum has
been calculated in \cite{article}. 
Due to the close similarities
between $\eta$ and $K$ mesons (pseudoscalars, similar masses),
and going to the SU(3) symmetric limit, we merely need to replace
$B\to K$ by $B \to \eta$ form factors and to take CKM factors
into account. At present, no reliable form factor calculations
for $B_{d,s}\to\eta$ exist. 

In the following, we estimate the
form factors for $B_s\to\eta$ by SU(3) relation to the $B\to
K$ form factors presented in \cite{article}, effectively
resulting in a multiplication of the form factors by $S_\eta$
(see eq.\ (\ref{eq:etaparm})). The calculation of $B_d\to\eta$
form factors is essentially identical, and so we omit an explicit
calculation.
\begin{eqnarray}
\braket{\eta(p_\eta)}{\bar{s}\gamma_\mu b}{B_s(p_B)} & = & \nonumber \\
& & \hspace{-30mm}S_\eta [ f_+^K(s)(p_B+p_\eta)_\mu + f_-^K(s)(p_B -
p_\eta)_\mu] \nonumber \\
\braket{\eta(p_\eta)}{\bar{s} \sigma_{\mu\nu}q^\nu b}{B_s(p_B)} & = & \nonumber \\
& & \hspace{-30mm}S_\eta i [ s(p_B+p_\eta)_\mu - (m_B^2 -
m_\eta^2)q_\mu]\textstyle{\frac{f_T^K(s)}{m_B+m_\eta}}
\end{eqnarray}
With these form factors, the total branching ratios for
\Bdsetallp\ as a function of $\theta_P$ has been calculated using
eq.~(\ref{eq:BKllpdiffbran}) integrated over the kinematical
region without intermediate $\psi$ resonant states included in
$C_9^{\mbox{eff}}$. 
\TABLE[ht]{\vspace*{3mm}\ \\
\begin{tabular}{|l|r|r|} \hline
    & $\theta_P = -10^\circ$ & $\theta_P = -20^\circ$
    \\ \hline $B_s \to \eta\!\ \mu^+\mu^-$ & $3.0\times 10^{-7}$
& $2.0\times 10^{-7}$
    \\ \hline  $B_d \to \eta\!\ \mu^+\mu^-$ & $7.7\times 10^{-9}$
& $1.0 \times 10^{-8}$\\ \hline
\end{tabular}
\caption{Branching ratios for \Bdsetallp.\label{tab:BRBetallp}}\vspace*{-3mm}}
\noindent A conservative theoretical uncertainty on this
calculation is given by $\approx 40\%$ from SU(3) breaking
corrections, and $\approx 25\%$ from uncertainties on the initial
$B\to K$ form factors, yielding a total uncertainty around 50\%.
Results are listed in table \ref{tab:BRBetallp} for the case $\ell = \mu$. 
Based on simulations from the two experiments of $B\to K^*\ell^+\ell^-$,
the number of reconstructed \Bsetammp\ and \Bdetammp\ events have been
estimated, again without taking the $\eta$ reconstruction efficiency into
account (table \ref{tab:Betallprec}).
\TABLE[h]{
\small{\begin{tabular}{|lr|r|r|}
    \hline
    & & \small{$B_{d} \to \eta\!\ \mu^+\mu^-$} &
    \small{$B_{s} \to \eta\!\ \mu^+\mu^-$}
\\ \hline
     CDF-II
       & $\theta_P=-10^\circ$
         & $8$
           & $60$
\\   (2 $\mbox{fb}^{-1}$)
       & $\theta_P=-20^\circ$
         & \small{$10$}
           & \small{$40$}
\\ \hline
     ATLAS
       & $\theta_P=-10^\circ$
         & \small{$20$}
           & \small{$160$}
\\   (30 $\mbox{fb}^{-1}$)
       & $\theta_P=-20^\circ$
         & \small{$25$}
           &\small{$105$}
    \\ \hline
     \multicolumn{4}{|c|}{\small $\eta$ not reconstructed ($\to$ factor
    $10\%-20\%$)}\\ \hline
    \end{tabular}}
\caption{Estimated number of reconstructed \Bdsetallp\ events at CDF-II and ATLAS.\label{tab:Betallprec}}
}

\section{Constraining $C_{10A}$ in semileptonic B decays}
\label{sec:constrainc10} The differential branching ratios of
$B_{d,s} \to K\ell^+\ell^-$ and $B_{s,d} \to \eta \ell^+\ell^-$
can be written as \cite{article}:
\begin{eqnarray}
\frac{d\Gamma}{d\hat{s}} & = & \frac{G_F^2\alpha^2m_B^5}{2^{10}\pi^5}\vert
V_{ts,td}^*V_{tb} \vert^2 \hat{u}(\hat{s})\left[ (\vert A'\vert^2+\vert
  C'\vert^2)(\lambda-\frac{\hat{u}(\hat{s})^2}{3})\right. \nonumber\\
& & + \vert C'\vert^2 4 \hat{m}_{\ell}^2(2+2\hat{m}_K^2-\hat{s}) +
  \mbox{Re}(C'{D'}^*) 8 \hat{m}_\ell^2(1-\hat{m}_K^2)\nonumber\\ & &\left. +\vert D'\vert^2 4
  \hat{m}_\ell^2\hat{s}\right]
\label{eq:BKllpdiffbran}
\end{eqnarray}
with
\begin{eqnarray}
\lambda_{K,\eta} \equiv \lambda(1,\hat{m}_{K,\eta}^2,\hat{s}) &
\hspace*{5mm} & \hat{u}(\hat{s}) \equiv \sqrt{\lambda\ \!
(1-4\frac{\hat{m}_\ell^2}{\hat{s}})}
\end{eqnarray}
and the coefficients $A'$, $C'$, $D'$ given in terms of Wilson
  coefficients and form factors as \cite{article}:
\begin{equation}
  A'(s) = C_9^{\mbox{eff}}(s) f_+^{K,\eta}(s) + \frac{2m_b}{m_B+m_{K,\eta}}
  C_7^{\mbox{eff}} f_T^{K,\eta}(s)
\end{equation}
\begin{eqnarray}
  C'(s) = C_{10A} f_+^{K,\eta}(s), & \hspace{10mm}&
  D'(s) = C_{10A} f_-^{K,\eta}(s).
\end{eqnarray}
Variables with \ $\hat{ }$ \ have been normalized by $m_B^2$,
$\lambda$ is the standard kinematical function, and $s$ is the
energy in the CM of the lepton pair. The $C_i$ entering these
expressions are the Wilson coefficients of the OPE operators
contributing to semileptonic $B$ decays:
\begin{equation}
\mathcal{O}_{7\gamma} =
\frac{e}{8\pi^2}m_b\bar{s}_\alpha\sigma^{\mu\nu}(1+\gamma_5)b^\alpha
F_{\mu\nu} \hspace*{0.7cm} \mathcal{O}^{\ell}_{9V} =
(\bar{s}_\alpha b^\alpha)_{V\! -\! A}(\bar{\ell}\ell)_{V}
\hspace*{0.7cm} \mathcal{O}^{\ell}_{10A} = (\bar{s}_\alpha
b^\alpha)_{V\! -\! A}(\bar{\ell}\ell)_{A}
\end{equation}
$C_7^{\mbox{eff}} = C_7+C_5/3-C_6$ and $C_9^{\mbox{eff}}$
contains corrections due to intermediate quark loops in the decay
(see e.g.~\cite{article}).

For zero lepton mass ($\ell=e,\mu$) only the $(|A'|^2 + |C'|^2)$ term in
(\ref{eq:BKllpdiffbran})
remains. Measuring $\mbox{d}\Gamma/d\hat{s}$ for e.g.\ the $B\to K
\mu^+\mu^-$ thus yields a way of experimentally determining this sum. This is
attractive since $A'$ is excessively difficult to
calculate theoretically, due to large hadronic uncertainties caused by the
existence of intermediate $\psi$ resonances entering $C_9^{\mbox{eff}}$.

In contrast to this,
the terms in (\ref{eq:BKllpdiffbran}) containing $C'$ and $D'$ depend only on
$C_{10A}$ and form factors. Eliminating $A'$ from the $\tau$ distribution
by isolating the term $(\vert A'\vert^2+\vert C'\vert^2)$ in the muon
distribution and inserting it in eq.\ (\ref{eq:BKllpdiffbran}) for
$\ell=\tau$ yields the following relation\footnote{The
result for $\Bsetammp$ is completely analogous.}, where we
have integrated over the kinematical region of the $\tau$ spectrum:
\begin{eqnarray}
\int_{4 \hat{m}_\tau^2}^{(\hat{m}_B-\hat{m}_K)^2} \hspace{-2mm}
\frac{dB_\tau}{d\hat{s}}d\hat{s} & = &
\int_{4 \hat{m}_\tau^2}^{(\hat{m}_B-\hat{m}_K)^2} \hspace{-2mm}
\sqrt{1-\frac{4\hat{m}_\tau^2}{\hat{s}}}\frac{dB_\mu}{d\hat{s}}\left(1+\frac{2
    \hat{m}_\tau^2}{\hat{s}}\right) d\hat{s} \nonumber \\
& & +
\vert C_{10A}\vert^2 \frac{G_F^2\alpha^2m_B^5}{2^{10}\pi^5}\vert
V_{ts}^*V_{tb} \vert^2 \tau_B \int_{4
  \hat{m}_\tau^2}^{(\hat{m}_B-\hat{m}_K)^2} \hspace{-2.5mm}F(\hat{s}) d\hat{s}
\label{theeq}
\end{eqnarray}
where $\tau_B$ is the $B$ meson lifetime and
\begin{eqnarray*}\textstyle{
F(\hat{s})} & = & \textstyle{\sqrt{\lambda
  (1-\frac{4\hat{m}_\tau^2}{\hat{s}})}\left(
  f_+^2 4 \hat{m}_\tau^2(2+2\hat{m}_K^2-\hat{s}) \right.}
  \textstyle{  \left. + f_+f_-8\hat{m}_\tau^2(1-\hat{m}_K^2)+f_-^24\hat{m}_\tau^2\hat{s}\right)}
  \nonumber
\end{eqnarray*}
To illustrate the order of magnitude of the terms entering eq.\
(\ref{theeq}), we give their SM values below (with the form
factors presented in \cite{article}).
\begin{eqnarray}
\int_{4 \hat{m}_\tau^2}^{(\hat{m}_B-\hat{m}_K)^2} \hspace{-2mm}
\frac{dB_\tau}{d\hat{s}}d\hat{s} & = & 1.35 \times 10^{-7} \\
\int_{4 \hat{m}_\tau^2}^{(\hat{m}_B-\hat{m}_K)^2}\! \! \! \! \! \! \!
\!\!\!\!\!\!\!\!\!\!\! d\hat{s} \ \ \
\sqrt{1-4\hat{m}_\tau^2/\hat{s}}\ \frac{dB_\mu}{d\hat{s}}\ (1 +
2\hat{m}_\tau^2/\hat{s}) & = & 0.90 \times 10^{-7} \\
|C_{10A}|^2\frac{G_F^2\alpha^2\hat{m}_B^5}{2^{10}\pi^5}\vert
V_{ts}^*V_{tb} \vert^2 \tau_B \int_{4
  \hat{m}_\tau^2}^{(\hat{m}_B-\hat{m}_K)^2} \! \! \! \! \! \! \!
\!\!\!\!\!\!\!\!\!\!\! d\hat{s} \ \ \ \ F(\hat{s}) &= & 0.45
\times 10^{-7}
\end{eqnarray}
Due to the close similarity between $K$ and $\eta$, these results are
not changed substantially when considering \Bsetammp. Using very precise form
factor calculations,
$F(s)$ can be evaluated theoretically to within perhaps $\pm 20\%$
uncertainty. At second generation machines, it is reasonable to expect
measurements of $BR(B\to K\tau^+\tau^-)$ and the muon spectrum to about
$10\%$ accuracy. Isolating $C_{10A}$ in the above formula would
then yield
an overall uncertainty of approximately $\pm 20 \%$. In the SUGRA models
investigated in \cite{goto98} (minimal and non-minimal, $\tan\beta=2$ and
$\tan\beta=30$), $C_{10A}$ lies within $\pm 10\%$ of the SM value, and so a
distinction is not yet possible for these cases. In the 2HDM\footnote{Two
Higgs Doublet Model}, $C_{10A}$ receives contributions from charged Higgs
bosons, and we have used
\cite{Aliev00} for values of the
Yukawa coupling of the charged Higgs to the top quark, $0 < \lambda_{tt} <
0.3$ and charged Higgs masses $0.2$ TeV $ < M_{H^\pm} < 1$ TeV. The case of
maximal deviation ($M_{H^\pm} = 1$ TeV, $\lambda_{tt}=0.3$) produces only
about 5\% deviation. Thus we see that some of the most common models
cannot be ruled out by a measurement such as the one proposed here in the
near future. Both form factor calculations and experimental precision must be
improved before this method becomes viable. Again, this result is not
substantially altered by having a $B_S$ initial state and an $\eta$ in the final state.


\section{Conclusion}
Using the method of quark topologies and invoking
SU(3)-suppression of certain topologies as well as SU(3) symmetry
of the strong interaction dynamics, it is possible to obtain a
parametrization of the \BdsJpsi\ amplitude which allows the
extraction of the angle $\gamma$ of the Unitarity Triangle through
measurement of $CP$ asymmetries in the \BdJpsi\ process and
measurement of the $CP$ averaged widths of the \BdsJpsi\
processes together. Electroweak penguins cannot lead to any
problems in this extraction. The measurement of asymmetries
require tagging whereas the $CP$ averaged rates do not.

Additionally, an estimate of the branching ratios for \BdJpsi\ has been
presented, yielding an expected branching ratio of around $1\times 10^{-5}$
for \BdJpsi\ and around $5\times 10^{-4}$ for \BsJpsi\ (taking $\theta_P
\approx -20^\circ$). Due to the difference in production ratios of $B_d$ and
$B_s$, this means a factor 5-10 more $B_s$ decays than $B_d$ decays in the
experiment, depending on the value of $\theta_P$. This is an improved
situation relative to the
$\BdsJpsiK_S$ decays, where a factor of around 250 difference is expected. It
has been estimated that HERA-B will probably not be able to see $\BdJpsi$
decays whereas the Tevatron experiments CDF-II/D0 perhaps stand a
chance of extracting of $\gamma$ from \BdsJpsi. Finally, the ATLAS
experiment should have sufficient statistics to allow the extraction of
$\gamma$ even taking the most pessimistic approach to the uncertainties in
the estimates of branching ratios and production rates presented in this
paper.

As an addendum, a strategy to constrain $C_{10A}$ in semileptonic $B$ decays
with small hadronic uncertainties
has been proposed. It has been found to require both high theoretical and
experimental precision, of a kind that will not be available in the near
future.

\label{sec:conclall}
\acknowledgments My greatest thanks to Alexander Khodjamirian for discussions
  and guidance. Many thanks also to Paula Eerola and to the HEP group at NBI
  for their
  help and support.
\bibliographystyle{unsrt}
\bibliography{JHEP_B}
\end{document}